\documentclass[9pt, sigconf]{acmart}

\AtBeginDocument{%
  }
 
\setcopyright{none}
\acmDOI{}
\acmISBN{}
\acmBooktitle{}
\acmConference[ISLPED '26]{International Symposium on Low Power Electronics and Design}{August 5 - 7, 2026}{Evanston, Illinois, USA}

\usepackage{pgfplots}
\usepackage{algorithm}
\usepackage{algorithmic}
\usepackage{graphicx}
\usepackage{textcomp}
\usepackage{xcolor}

\usepackage{booktabs} 
\usepackage[caption=false]{subfig}
\usepackage{fancyvrb} 
\usepackage{multirow}
\usepackage{diagbox}
\usepackage{tikz}
\usepackage{numprint}
\usepackage{url}
\usepackage{colortbl}
\usepackage{dhucs} 

\usepackage{enumitem} 
\usepackage{makecell} 

\usepackage[font=small]{caption}
\usepackage{comment}
\usepackage{lineno}
\usepackage{color} 
\usepackage{setspace} 
\usepackage{relsize}
\usepackage{tabularx}
\usepackage{bm}
\usepackage{hhline}
\usepackage{xurl}

\usepackage{ragged2e}
\usepackage{stfloats}

\setcitestyle{numbers,sort&compress}

\newcommand\refFigure[1]{Figure~\ref{#1}}
\newcommand\refTable[1]{Table~\ref{#1}}
\newcommand\refSection[1]{Section~\ref{#1}}

\newcommand\korean[1]{}
\newcommand\blind[1]{XXXX}

\renewcommand{\baselinestretch}{1}

\begin{document}
\title[CMAX-CAMEL]{{CMAX-CAMEL: A Coarse-to-Fine Adaptive, Memory-Efficient, and Low-Power Edge Processor for Contrast Maximization}}

\author{Kyeongpil Min, Jongin Choi, Kyeongwon Lee, and Woojoo Lee}
\affiliation{%
  \institution{
}
  \city{}
  \country{} 
}
\email{}
\thanks{This paper has been accepted for publication in the Proceedings of the ACM/IEEE International Symposium on Low Power Electronics and Design (ISLPED), 2026.}

\thanks{
This work was supported in part by the National Research Foundation of Korea (NRF) grant funded by MSIT (No. RS-2024-00345668), and in part by Institute of Information \& communications Technology Planning \& Evaluation (IITP) grants funded by MSIT (No. RS-2023-00277060).

Kyeongpil Min and Jongin Choi contributed equally to this work. 

Woojoo Lee is the corresponding author.}

\begin{abstract}
Contrast maximization (CMAX) is a direct geometric framework for event-based motion estimation, but its iterative warp-and-accumulate pipeline incurs input-dependent computation and frequent memory accesses, challenging real-time, low-power edge deployment. We present CMAX-CAMEL, a coarse-to-fine adaptive, memory-efficient, low-power edge processor for CMAX. CMAX-CAMEL combines a runtime-adaptive execution strategy with a memory-centric processor architecture. It adjusts coarse-to-fine execution according to the observed event distribution, prioritizing stages likely to improve estimation accuracy while suppressing low-value iterations and unnecessary stage transitions. Architecturally, a banked parallel memory organization sustains real-time throughput while reducing latency, and a subsampling-coupled accumulation structure lowers memory-access activity along the warp-and-accumulate dataflow. On a Virtex FPGA prototype operating at 200\,MHz, CMAX-CAMEL improves estimation accuracy by up to 19\% over fixed coarse-to-fine schedules, reduces processing latency by 53.3\%, lowers effective memory accesses by 42\%, and cuts total system energy by 52.2\%, including adaptation overheads. These results show that CMAX-CAMEL is an HW--SW co-design that co-optimizes execution policy and data movement for real-time, low-power event-based motion estimation at the edge.
\end{abstract}

\maketitle

\section{Introduction}\label{sec:intro}

Autonomous and mobile edge systems require motion estimation that remains reliable under fast motion, abrupt illumination changes, and stringent energy and latency constraints~\cite{gehrig:nature2024,kuhne:sensors2024}
This capability is a fundamental front-end component of visual SLAM and odometry, where the ego-motion of a camera must be inferred from sequential observations~\cite{wang:TIM2024,xu:robotics2025}. 
Event cameras, especially Dynamic Vision Sensors (DVS)~\cite{lichtsteiner:JSSC2008}, are well suited to this setting because they asynchronously report local brightness changes~\cite{Rebecq:LRA17, CHOI:AEJ25, Cimarelli:sensors25}.
They offer microsecond-level temporal resolution and wide dynamic range for handling fast motion and challenging lighting, while generating sparse outputs that reduce data movement~\cite{aliakbarpour:Sensors2024,gallego:TPAMI2020}.
Under the resource and power constraints of edge deployment, however, pure event-based pipelines remain particularly attractive because they operate directly on the event stream without relying on auxiliary sensing modalities or power-hungry inference~\cite{Kueng:IROS16, chamorro:robotics2022,zhou:robotics2021}. 
Within this class of methods, contrast maximization (CMAX)~\cite{Gallego:LRA17}, which estimates motion by warping events to maximize the contrast of an accumulated event image, has emerged as a representative geometric framework and has been extended to rotational motion estimation, SLAM, and motion segmentation~\cite{gallego:CVPR2018, Kim:LRA21, Zhou:TNNLS23, Shiba:TPAMI24,guo:Robotics2024,stoffregen:ICCV2019, Yamaki:CVPRW25}.
Yet, despite substantial algorithmic progress, processor- and system-level support needed to make CMAX a practical real-time, low-power primitive for edge platforms remains limited.

Recent efforts have explored both coarse-to-fine execution and hardware acceleration for CMAX~\cite{wang2025frme,min2026coarse}. Collectively, these studies suggest two key design opportunities: the utility of CMAX computation is highly non-uniform across stages and iterations, and the repeated warp-and-accumulate dataflow can benefit substantially from dedicated architectural support. 
However, neither direction fully addresses the needs of always-on edge sensing. 
Coarse-to-fine methods generally rely on predetermined schedules rather than adapting computation to runtime event characteristics, which can vary significantly over time. Existing accelerators, in turn, largely optimize throughput within a fixed iterative flow and do not directly reduce the memory-access activity that often dominates system energy in CMAX pipelines. 
For edge deployment, reducing arithmetic cost alone is therefore insufficient unless the associated data movement and latency are co-optimized as well~\cite{silvano:survey2025,akkad:TAI2023}.

These limitations reveal a fundamental co-design challenge: in a resource-constrained edge system, the goal is not merely to accelerate a fixed workload, but to adapt the workload itself to the incoming event distribution~\cite{golpayegani:TAAS2024,tuli:JNPA2023}.
The execution policy should prioritize computations that materially improve the motion estimate while suppressing low-value iterations and unnecessary stage transitions. 
At the same time, the hardware should reduce memory-access activity along the warping and accumulation path without sacrificing real-time throughput. These objectives are tightly coupled and jointly determine the achievable accuracy--energy--latency trade-off of edge CMAX processing. 
Nevertheless, prior work has not unified runtime-adaptive execution and memory-centric processor architecture within a single HW--SW co-design framework.

To address this gap, we present \textbf{CMAX-CAMEL}, a \textbf{C}oarse-to-fine \textbf{A}daptive, \textbf{M}emory-\textbf{E}fficient, \textbf{L}ow-power edge processor for contrast maximization in event-based motion estimation.
CMAX-CAMEL combines runtime-adaptive CMAX execution with a memory-centric processor architecture tailored to the warp-and-accumulate dataflow. The main contributions of this work are as follows:

\begin{enumerate}[leftmargin=*,nosep,label=\textbf{\arabic*)}]
    \item \textbf{Runtime-adaptive CMAX execution.}
    Rather than following a fixed coarse-to-fine schedule, CMAX-CAMEL dynamically adjusts its execution configuration according to the observed event distribution at runtime. This preserves the efficiency benefits of coarse-to-fine execution while preferentially allocating compute resources to stages that are more likely to improve estimation accuracy, thereby suppressing low-value iterations and unnecessary stage transitions.

    \item \textbf{CMAX-CAMEL engine architecture.}
    We propose a memory-centric processor architecture that targets memory-access activity, a major contributor to energy in edge CMAX processing. A banked parallel memory organization sustains real-time throughput and reduces processing latency, while a subsampling-coupled accumulation structure reduces effective memory accesses during coarse-to-fine execution. The complete design is prototyped and validated on a Virtex FPGA prototype operating at 200\,MHz.
\end{enumerate}

Experimental results on representative target scenarios show that the proposed runtime-adaptive execution strategy improves estimation accuracy by up to 19\% over fixed coarse-to-fine schedules while retaining the efficiency benefits of coarse-to-fine execution. 
In addition, CMAX-CAMEL reduces processing latency by 53.3\%, lowers effective memory accesses by 42\%, and reduces total system energy by 52.2\%, even after accounting for adaptation overheads. 
These results show that CMAX-CAMEL is not merely a faster implementation of CMAX, but a low-power edge processor that co-optimizes execution policy and data movement for real-time event-based motion estimation.

\section{Background and Motivation}\label{sec:background}

\begin{figure}[t]
\centering
\includegraphics[width=0.9\columnwidth]{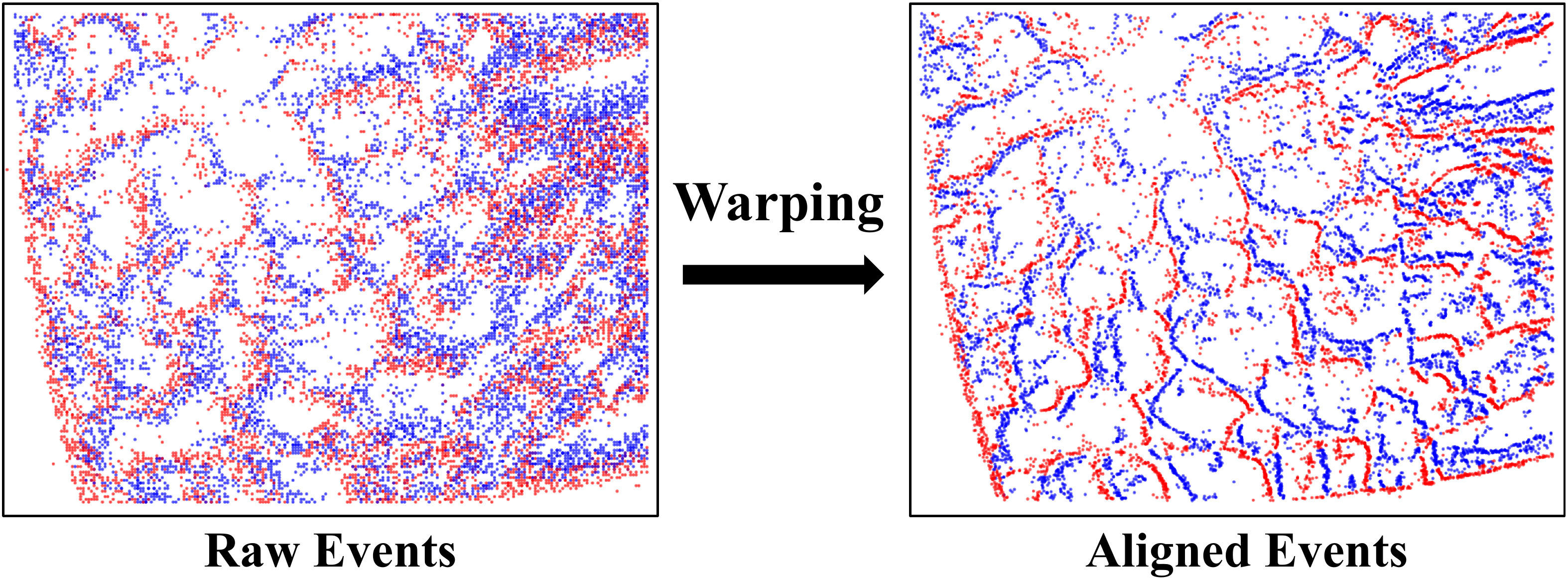}
\vskip -10pt
\caption{CMAX warps scattered raw events with estimated motion parameters to form a sharp image of warped events (IWE) aligned along scene edges. Red/blue: positive/negative polarity.}
\label{fig:intro}
\end{figure}

\noindent \textit{\textbf{Overview of CMAX.}}
CMAX is a geometric framework for event-based ego-motion estimation that relies on the observation that, under correct motion compensation, events generated by the same physical edge become more tightly aligned on the image plane, thereby increasing the spatial contrast of the accumulated event image. 
Each event asynchronously generated by a DVS when the brightness change at a pixel exceeds a threshold can be represented as $e_k=(\mathbf{x}_k,t_k,p_k)$, where $\mathbf{x}_k=(x_k,y_k)$ denotes the pixel location, $t_k$ the timestamp, and $p_k\in\{-1,+1\}$ the polarity. 
An event window of $N$ events is denoted by $\mathcal{E}=\{e_k\}_{k=1}^{N}$. 
CMAX estimates the motion parameters that best align the events within a window, i.e., those that maximize the contrast of the motion-compensated event image. 
Under the pure-rotation assumption, an event observed at time $t_k$ is warped to a common reference time $t_0$ under a motion hypothesis $\boldsymbol{\omega}$:
{\small
\begin{equation}
\mathbf{x}'_k = W(\mathbf{x}_k;\boldsymbol{\omega}, t_k-t_0).
\end{equation}}%
The warped events are then accumulated into an image of warped events (IWE), as illustrated in \refFigure{fig:intro},
{\small\begin{equation}
I(\mathbf{x};\boldsymbol{\omega})=\sum_{k=1}^{N} p_k \, \delta\!\left(\mathbf{x}-\mathbf{x}'_k(\boldsymbol{\omega})\right),
\end{equation}}%
where $\delta(\cdot)$ denotes the Dirac delta~\cite{gallego:CVPR2018}. 
In practice, this accumulation is implemented on a discrete pixel grid using bilinear voting~\cite{gallego:CVPR2018}. 
Because the event stream is sparse, the resulting objective surface can be highly non-smooth. 
To alleviate this issue, Gaussian smoothing~\cite{gallego:CVPR2019} is applied:
{\small\begin{equation}
I_{\sigma}(\mathbf{x};\boldsymbol{\omega}) = \left(I(\cdot;\boldsymbol{\omega}) * G_{\sigma}\right)(\mathbf{x}).
\end{equation}}%
The alignment quality is then measured by the spatial contrast of the smoothed IWE, typically its variance,
{\small\begin{equation}
C(\boldsymbol{\omega}) = \mathrm{Var}(I_{\sigma}(\mathbf{x};\boldsymbol{\omega})),
\end{equation}}%
which increases as the events become better aligned. The optimal motion hypothesis is therefore given by
{\small\begin{equation}
\boldsymbol{\omega}^{*} = \arg\max_{\boldsymbol{\omega}} C(\boldsymbol{\omega}).
\end{equation}}%
In practice, CMAX solves this problem by gradient-based iterative optimization using $\nabla_{\boldsymbol{\omega}} C(\boldsymbol{\omega})$. 
For consecutive event windows, the estimate from the previous window is typically reused as a warm-start~\cite{Gallego:LRA17} initialization to improve convergence and reduce the number of iterations. 
However, conventional CMAX repeats event warping, IWE accumulation, Gaussian smoothing, and contrast/gradient evaluation over the same event set and the full-resolution IWE at every iteration, resulting in an event-domain cost proportional to the number of events and an image-domain cost proportional to the IWE resolution.

\begin{figure}[t]
\centering
\includegraphics[width=\columnwidth]{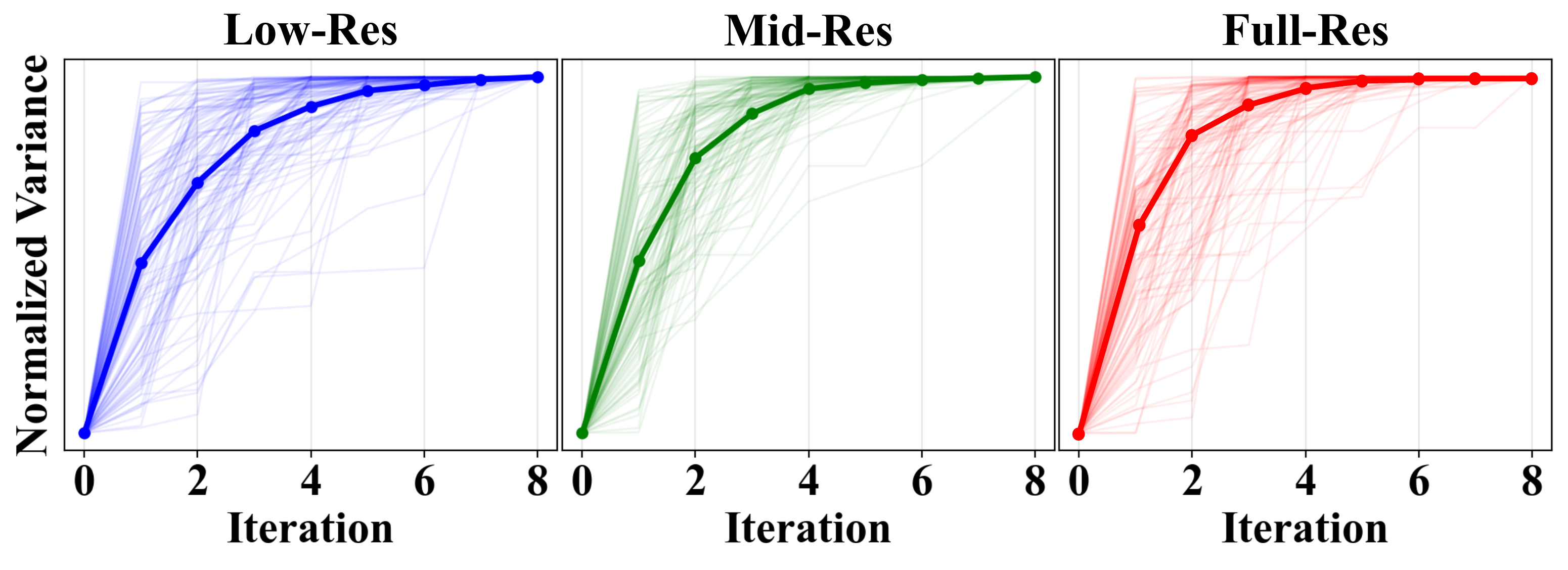}
\vskip -6pt
\caption{Empirical stage-wise convergence behavior of coarse-to-fine CMAX across the Low-Resolution, Mid-Resolution, and Full-Resolution stages. In each stage, the normalized variance rises rapidly at first and then gradually saturates. Thin lines show individual event windows, and bold lines show the mean.}
\label{fig:iteration}
\end{figure}

\noindent \textit{\textbf{Coarse-to-Fine Strategy.}}
Empirically, the convergence of CMAX exhibits a coarse-to-fine pattern: early iterations mainly correct large misalignment, whereas later iterations provide progressively finer refinement and smaller contrast gains, as shown in \refFigure{fig:iteration}. 
Conventional CMAX does not exploit this non-uniform convergence in its execution structure. 
A recent coarse-to-fine formulation addresses this mismatch by jointly scaling the IWE resolution and the event workload across stages~\cite{min2026coarse}. 
Let $\mathcal{E}_s \subseteq \mathcal{E}$ denote the event subset retained at resolution scale $s$. The scaled IWE is then
{\small\begin{equation}
I_s(\mathbf{u};\boldsymbol{\omega}) =
\sum_{e_k \in \mathcal{E}_s} p_k \, K\!\left(\mathbf{u} - s\,\mathbf{x}'_k(\boldsymbol{\omega})\right),
\end{equation}}%
where $K(\cdot)$ denotes the bilinear voting kernel used to realize the accumulation on a discrete grid. 
If the full-resolution IWE has height $H$ and width $W$, the scaled grid becomes $H_s=\lceil sH\rceil$ and $W_s=\lceil sW\rceil$, so the image-domain workload decreases approximately with $s^2$.
In addition, coarse stages apply event subsampling to suppress redundant contributions from events that are warped into the same coarse grid cell. 
The keep ratio is set to $\rho_s=s$, so that early stages use fewer events for low-cost coarse alignment, while finer stages progressively restore more events for refinement. 
Since this subsampling is performed only once using a warm-start-based reference warp at the beginning of each stage, its additional overhead remains modest. 
Overall, the coarse-to-fine strategy reduces the total computation of iterative CMAX by jointly controlling IWE resolution and event count across stages. 
However, its stage schedule remains predetermined rather than runtime-adaptive.


\noindent \textit{\textbf{System Design Motivation.}}
CMAX provides a direct geometric foundation for event-based motion estimation, and the coarse-to-fine strategy shows that the repeated computation in its iterative optimization can be significantly reduced. 
However, fixed coarse-to-fine schedules are still insufficient for practical always-on edge systems. 
As shown in \refFigure{fig:iteration}, the mean normalized variance improvement within each stage tends to saturate as iterations proceed, suggesting that the iteration budget of a stage should be adjusted according to the observed convergence state rather than predetermined offline. 
At the same time, individual event windows exhibit substantial variation in both the rate and the saturation point of this improvement, indicating that stage residence should also depend on input-specific runtime dynamics. The problem therefore extends beyond repeatedly executing a fixed iterative flow; it becomes an execution-control problem that decides how computation should be allocated across stages based on runtime evidence. 
Importantly, this adaptive execution problem should be treated separately from the repeated warp-and-accumulate datapath itself. 
Such a separation allows the datapath to be optimized for low power in terms of memory-access activity and processing latency, while the control layer remains flexible enough to accommodate input-dependent execution policies. 
These observations motivate a system-level HW--SW co-design that couples runtime-adaptive execution control with a memory-centric processor architecture for real-time, low-power edge deployment of CMAX.


\begin{algorithm}[t]
\footnotesize
\caption{Runtime-Adaptive CMAX Stage Transition}
\label{alg:runtime_stage}
\begin{algorithmic}[1]
\STATE \textbf{Input:} initial estimate $\boldsymbol{\omega}$, stages $\mathcal{S}=\{s_{1/4},s_{1/2},s_{1}\}$, thresholds $\{\tau_{1/4},\tau_{1/2},\tau_{1}\}$
\STATE $s \gets s_{1/4}$; \quad $V^{\mathrm{prev}} \gets V_s(\boldsymbol{\omega})$
\WHILE{true}
    \STATE $\boldsymbol{\omega} \gets \mathrm{Update}(\boldsymbol{\omega}, s)$
    \STATE $V \gets V_s(\boldsymbol{\omega})$
    \STATE $g \gets \dfrac{V - V^{\mathrm{prev}}}{|V^{\mathrm{prev}}|}$ \hfill // normalized variance gain
    \IF{$g \ge \tau_s$}
        \STATE $V^{\mathrm{prev}} \gets V$ \hfill // keep the current stage
    \ELSIF{$s \neq s_{1}$}
        \STATE $s \gets \mathrm{Next}(s)$; \quad $V^{\mathrm{prev}} \gets V_s(\boldsymbol{\omega})$ \hfill // promote to the next finer stage
    \ELSE
        \STATE \textbf{break} \hfill // terminate at the finest stage
    \ENDIF
\ENDWHILE
\STATE \textbf{return} $\boldsymbol{\omega}$
\end{algorithmic}
\end{algorithm}

\section{Runtime-Adaptive CMAX Execution}\label{sec:method}

The main limitation of fixed coarse-to-fine CMAX lies not in the stage-wise motion-refinement operator itself, but in the static schedule used to control it. 
CMAX-CAMEL therefore retains the same coarse-to-fine stages and per-stage update kernels as the baseline, but replaces the fixed iteration budget with a runtime-adaptive stage-transition policy. 
Instead of allocating a predetermined number of iterations to each stage, the controller evaluates whether the current stage continues to provide meaningful alignment improvement for the current event window and then either remains at the current stage, advances to the next finer stage, or terminates at the finest stage. This preserves the computational structure of coarse-to-fine CMAX while adapting the allocation of limited compute resources to the input-dependent convergence behavior.


Because CMAX directly maximizes the spatial variance of the motion-compensated IWE, stage-wise variance provides a natural indicator of whether additional refinement at the current stage remains beneficial. Let $V_s^{(i)}$ denote the variance of the stage-$s$ IWE after the $i$-th update at that stage. We define the normalized variance gain as
{\small\begin{equation}
g_s^{(i)} =
\frac{V_s^{(i)} - V_s^{(i-1)}}{\left|V_s^{(i-1)}\right|},
\label{eq:stage_gain}
\end{equation}}%
where $s\in\left\{\frac{1}{4},\frac{1}{2},1\right\}$ is the 
resolution scale that identifies each stage (low-, mid-, and full-resolution respectively). If $g_s^{(i)} \ge \tau_s$, the current stage is retained because it is still yielding meaningful objective improvement. Otherwise, execution advances to the next finer stage; at the full-resolution stage, the same condition terminates the optimization. The threshold $\tau_s$ thus acts as a stage-specific sensitivity parameter that determines when stage-wise gain has saturated enough to justify progression.

Algorithm~\ref{alg:runtime_stage} summarizes the proposed policy. The algorithm starts from the coarsest stage using the warm-start estimate from the previous event window. 
After each update, it evaluates the normalized variance gain and compares it with the threshold of the current stage. 
Importantly, CMAX-CAMEL does not alter the underlying CMAX objective or the per-stage update rule; it only changes the high-level execution policy that controls stage residence and transition timing. Consequently, the method preserves the efficiency benefits of coarse-to-fine CMAX while mitigating the behavioral mismatch introduced by a fixed schedule. 
For the experiments reported here, the threshold values $\{\tau_{1/4},\tau_{1/2},\tau_{1}\}$, are selected empirically. More generally, however, the same policy form can accommodate different accuracy--efficiency operating points.

\begin{figure}[t]
\centering
\includegraphics[width=1\columnwidth]{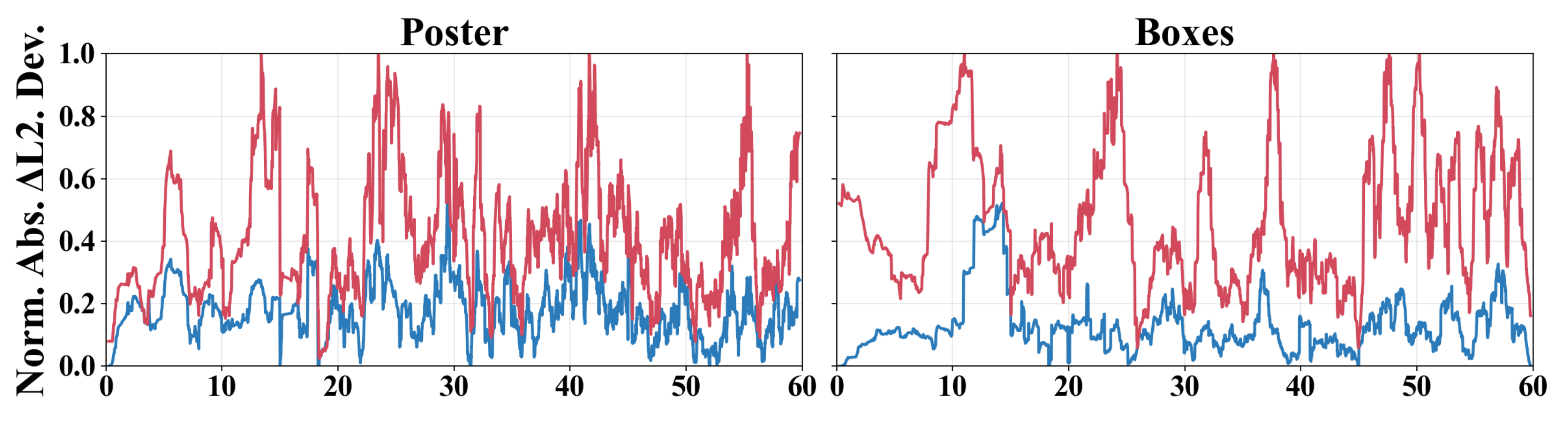}
\vskip -5pt
\caption{Normalized absolute L2 deviation from full-resolution CMAX on the \texttt{poster} and \texttt{boxes} sequences. Lower values indicate behavior closer to full-resolution CMAX. (Red: fixed-schedule CMAX; Blue: proposed runtime-adaptive CMAX.)}
\label{fig:fig3}
\end{figure}

\begin{table}[t]
\centering
\caption{IMU-referenced angular-velocity RMSE of full-resolution, fixed-schedule, and runtime-adaptive CMAX.}
\vskip -8pt
\small
  \renewcommand{\arraystretch}{1.05}
\begin{tabular}{c|c|c|c}
\Xhline{1pt}
\textbf{Sequence} &
\shortstack{\textbf{Full-resolution}\\\textbf{CMAX}} &
\shortstack{\textbf{Fixed-schedule}\\\textbf{CMAX}} &
\shortstack{\textbf{Proposed}\\\textbf{Method}} \\
\Xhline{0.8pt}
poster & 8.39 & 11.30 & 9.56 \\
boxes  & 5.39 & 7.04  & 5.65 \\
\Xhline{1pt}
\end{tabular}
\label{tab:rmse_comparison}
\end{table}

To quantify the benefit of runtime adaptivity, we evaluate the \texttt{poster} and \texttt{boxes} sequences from the Event Camera Dataset and Simulator~\cite{Mueggler:IJRR17} using inertial measurement unit (IMU) angular velocity as reference. We compare three methods: full-resolution CMAX (the conventional implementation without coarse-to-fine scaling), fixed-schedule coarse-to-fine CMAX, and the proposed runtime-adaptive CMAX. 
Because both reduced-cost variants aim to approximate the behavior of the full-resolution baseline at lower cost, we examine not only IMU-referenced estimation accuracy but also behavioral closeness to full-resolution CMAX. 
Let $\hat{\boldsymbol{\omega}}_m[k]$ denote the angular-velocity estimate of method $m$ for window $k$, and let $\boldsymbol{\omega}_{\mathrm{IMU}}[k]$ denote the reference angular velocity. We define
{\small
\begin{equation}
\label{eq:dev_metric}
\begin{aligned}
e_m[k] &= \left\|\hat{\boldsymbol{\omega}}_m[k] - \boldsymbol{\omega}_{\mathrm{IMU}}[k]\right\|_2,\\
D_m[k] &= \mathcal{N}\!\left(\left|e_m[k] - e_{\mathrm{full}}[k]\right|\right),
\end{aligned}
\end{equation}}%
where $e_m[k]$ is the IMU-referenced angular-velocity error and $D_m[k]$ is a normalized absolute deviation from the full-resolution CMAX baseline. Here, $\mathcal{N}(\cdot)$ denotes min--max normalization applied independently over four 15-s segments (0--15\,s, 15--30\,s, 30--45\,s, and 45--60\,s) to account for different motion regimes within a sequence. Lower $D_m[k]$ indicates behavior closer to full-resolution CMAX.

\refFigure{fig:fig3} and \refTable{tab:rmse_comparison} show that the proposed runtime-adaptive execution more closely tracks the full-resolution baseline than fixed-schedule coarse-to-fine CMAX, and that this behavioral closeness translates into improved estimation accuracy. Across both \texttt{poster} and \texttt{boxes}, the deviation from full-resolution CMAX is consistently lower for the proposed method. 
Correspondingly, the IMU-referenced RMSE decreases from 11.30 to 9.56 on \texttt{poster} and from 7.04 to 5.65 on \texttt{boxes}, corresponding to reductions of 15.4\% and 19.7\%, respectively, relative to fixed-schedule coarse-to-fine CMAX. 
These results indicate that runtime-adaptive stage control mitigates the mismatch of static schedules and more faithfully preserves the optimization behavior of full-resolution CMAX under limited computation. 
This runtime-adaptive execution policy provides the algorithmic foundation for the processor architecture described in \refSection{sec:hw}.

\section{CMAX-CAMEL Engine Architecture}\label{sec:hw}

\begin{figure*}[t]
\centering
\includegraphics[width=1.95\columnwidth]{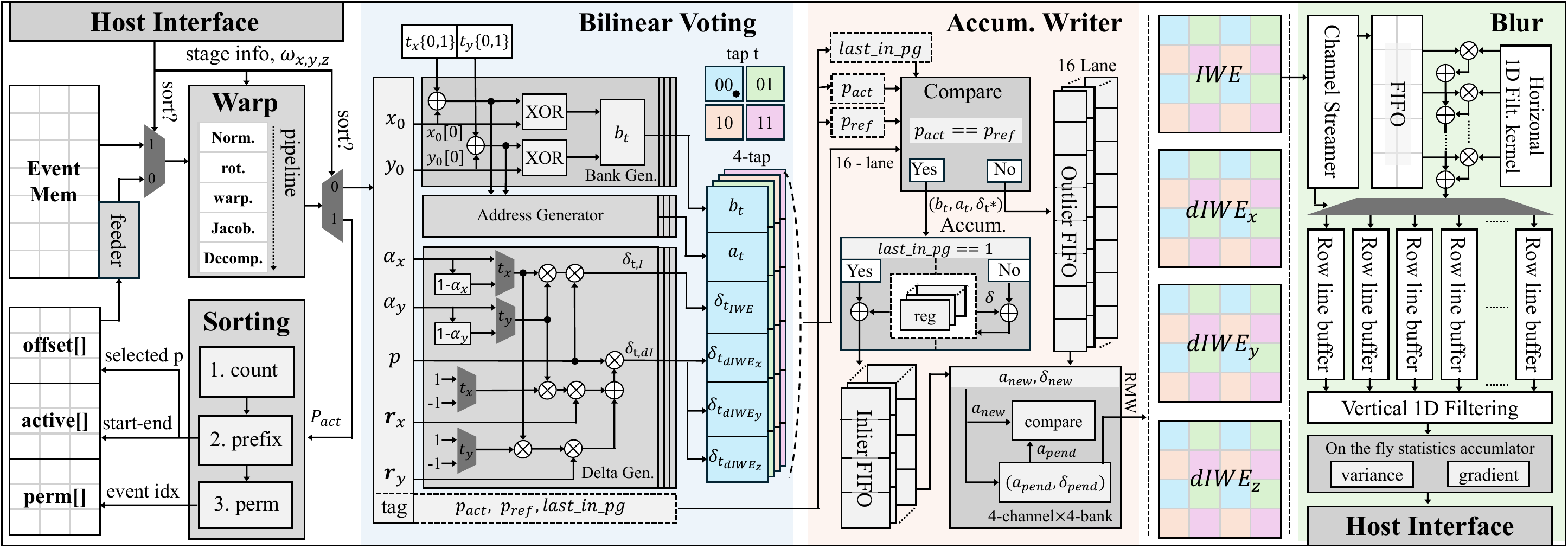}
\caption{Overall architecture of the CMAX-CAMEL engine. The controller determines stage transitions and initiates stage-local sorting, while the datapath executes each stage through pixel-grouped sorting with a shared warp front-end, conflict-free banked bilinear voting, local accumulation with pending merge, and streaming Gaussian smoothing with on-the-fly statistics.}
\label{fig:CAMEL}
\end{figure*}

\begin{algorithm}[t]
\footnotesize
\caption{Shared pipelined event warp}
\label{alg:warp}
\begin{algorithmic}[1]
\STATE \textbf{Input:} event $(x,y,t,p)$, reference time $t_{\mathrm{ref}}$, $\boldsymbol{\omega}=(\omega_x,\omega_y,\omega_z)$, scale $s$
\STATE \textbf{Output:} $(x_0,y_0)$, $(\alpha_x,\alpha_y)$, $(\mathbf{r}_x,\mathbf{r}_y)$, $p_{\mathrm{act}}$
\STATE $x_n\gets(x-c_x)/f_x$;\; $y_n\gets(y-c_y)/f_y$;\; $\Delta t\gets t-t_{\mathrm{ref}}$ \hfill // Normalizing
\STATE $B\gets 1+x_n^2$;\; $D\gets 1+y_n^2$;\; $XY\gets x_n y_n$ \hfill // Computing rotation terms
\STATE $u\gets f_x(XY\omega_x-B\omega_y+y_n\omega_z)$ \hfill 
\STATE $v\gets f_y(D\omega_x-XY\omega_y-x_n\omega_z)$ \hfill 
\STATE $(x',y')\gets s\cdot(x-\Delta t\cdot u,\; y-\Delta t\cdot v)$ \hfill // Scaling and warping
\STATE $\mathbf{r}_x\gets s\Delta t\cdot[f_xXY,\,-f_xB,\,f_xy_n]$ \hfill // Building Jacobian rows
\STATE $\mathbf{r}_y\gets s\Delta t\cdot[f_yD,\,-f_yXY,\,-f_yx_n]$ \hfill 
\STATE $x_0\gets\lfloor x'\rfloor$;\; $y_0\gets\lfloor y'\rfloor$ \hfill // Decomposing subpixel coordinates
\STATE $\alpha_x\gets x'-x_0$;\; $\alpha_y\gets y'-y_0$ \hfill
\IF{in range}
    \STATE $p_{\mathrm{act}}\gets y_0\cdot W_s + x_0$ \hfill  
\ELSE
    \STATE $p_{\mathrm{act}}\gets \texttt{invalid}$
\ENDIF
\end{algorithmic}
\end{algorithm}

Implementing the runtime-adaptive execution policy of \refSection{sec:method} on an edge platform requires not only stage-transition control but also a datapath that suppresses the dominant cost of repeated warp-and-accumulate operations. In each CMAX iteration, the processor must form the IWE together with three derivative images, $\mathrm{dIWE}_{x,y,z}$, where $\mathrm{dIWE}_{j}$ denotes the per-pixel sensitivity of the warped event image to $\omega_j$ for $j\in\{x,y,z\}$, then perform Gaussian smoothing and evaluate the contrast objective and its gradient. This flow exposes two bottlenecks: multi-tap bilinear updates can stall the pipeline, and event-wise read-modify-write updates cause excessive effective memory accesses that dominate both energy and latency.


The CMAX-CAMEL engine addresses these bottlenecks with four coordinated blocks: 1) pixel-grouped sorting, 2) conflict-free banked bilinear voting, 3) local accumulation with pending merge, and 4) streaming Gaussian smoothing with on-the-fly statistics. Figure~\ref{fig:CAMEL} shows how these blocks are organized into a unified stage-execution pipeline. At stage entry, the controller provides the current scale factor and motion estimate and triggers the sorting block to construct the stage-local metadata tables (\texttt{active}, \texttt{offset}, and \texttt{perm}). These tables are reused across all iterations of the stage to stream events in pixel-group order. The shared warp front-end computes the warped coordinates, bilinear-voting factors, and Jacobian terms required by the downstream datapath. The bilinear stage converts each event into banked multi-tap updates for the IWE and derivative-image channels, after which the accumulation block merges inlier updates locally and applies pending merge before memory commit. Finally, the accumulated IWE and $\mathrm{dIWE}_{x,y,z}$ channels are consumed by the streaming Gaussian-smoothing block, which updates the statistics needed for variance and gradient evaluation without writing back blurred images.

\noindent \textit{\textbf{Pixel-Grouped Sorting with Shared Warp Front-End.}}
To exploit locality and support stage-wise subsampling, the input event window is reorganized into pixel-group runs before accumulation. The shared warp front-end in Algorithm~\ref{alg:warp} computes the stage-scaled warped location of each event and outputs $(x_0,y_0)$, $(\alpha_x,\alpha_y)$, Jacobian rows $(\mathbf{r}_x,\mathbf{r}_y)$, and the stage-local index $p_{\mathrm{act}}$. The same front-end is reused later in the main datapath, reducing hardware cost.


\begin{algorithm}[t]
\footnotesize
\caption{Pixel-grouped sorting with stage-aware subsampling}
\label{alg:pixel_sort}
\begin{algorithmic}[1]
\STATE \textbf{Input:} events $\{e_i\}_{i=0}^{N-1}$, pixel count $P$, stage $s$
\STATE \textbf{Output:} active$[\,]$, offset$[\,]$, perm$[\,]$
\STATE cnt$[0..P\!-\!1]\gets 0$ \hfill // State 1: Counting events per pixel-group
\FOR{$i=0$ to $N\!-\!1$}
    \STATE gid$[i]\gets \textsc{Warp}(e_i).p_{\mathrm{act}}$
    \IF{valid(gid$[i]$)}
        \STATE cnt[gid$[i]$]$++$
    \ENDIF
\ENDFOR
\STATE sum$\gets 0$;\; $M\gets 0$ \hfill // State 2: Building offsets, applying stage policy
\FOR{$p=0$ to $P\!-\!1$}
    \STATE offset$[p]\gets$ sum
    \STATE $k$, stride$[p]$, act$[p]\gets \textsc{StagePolicy}(\text{cnt}[p],s)$
    \IF{act$[p]$}
        \STATE active$[M]\gets p$;\; $M\!+\!+$;\; sum$\gets$ sum $+$ $k$
    \ENDIF
\ENDFOR
\STATE offset$[P]\gets$ sum;\; ptr$\gets$ offset;\; rank$[0..P\!-\!1]\gets 0$ \hfill // State 3: Permuting retained events
\FOR{$i=0$ to $N\!-\!1$}
    \STATE $p\gets$ gid$[i]$
    \IF{valid($p$) \AND act$[p]$}
        \IF{$(\text{rank}[p]\bmod \text{stride}[p]=0)$ \AND $(\text{ptr}[p]<\text{offset}[p+1])$}
            \STATE perm$[\text{ptr}[p]]\gets i$;\; ptr$[p]\!+\!+$
        \ENDIF
        \STATE rank$[p]\!+\!+$
    \ENDIF
\ENDFOR
\end{algorithmic}
\end{algorithm}

Algorithm~\ref{alg:pixel_sort} summarizes the 3-state sorting flow. In the count state, valid warped events are counted per stage-local group. In the prefix-scan state, \texttt{offset[]} is generated, and the stage policy is applied to determine the retained-event budget, subsampling stride, and activity flag of each group. In the perm state, the retained event indices are permuted into pixel-group order. Subsampling is applied using the group-local rank rather than the global event index, so the retention policy is enforced independently for each group. Since warm-start initialization keeps inter-iteration motion updates local, sorting is performed once at stage entry and reused across all iterations of that stage. The resulting \texttt{active[]}, \texttt{offset[]}, and \texttt{perm[]} tables allow the feeder to stream each run together with $p_{\mathrm{ref}}$ and \texttt{last\_in\_pg} for later accumulation.

\noindent \textit{\textbf{Conflict-Free Banked Bilinear Voting.}}
After sorting, the reordered event stream passes through the warp front-end and a bilinear voting stage that generates updates to the IWE and derivative-image memories. For each event, bilinear voting distributes the contribution to the four neighbors of $(x_0,y_0)$. Across the four channels $(\mathrm{IWE}, \mathrm{dIWE}_{x}, \mathrm{dIWE}_{y}, \mathrm{dIWE}_{z})$, a naive design would therefore trigger up to $16$ event-wise updates, creating frequent bank conflicts and pipeline stalls.

\begin{table}[t]
\centering
\caption{Stage-wise locality statistics on the \texttt{poster} sequence, normalized to the input event count.}
\vskip -8pt
\small
  \renewcommand{\arraystretch}{1.05}
\begin{tabular}{c|c|c|c}
\Xhline{1pt}
\textbf{Metric (\% of events)} &
\textbf{Low-Res} &
\textbf{Mid-Res} &
\textbf{Full-Res} \\
\Xhline{0.8pt}
Active pixel-group ratio & 21.5 & 34.1 & 59.1 \\
Outlier ratio            & 5.2  & 4.7  & 5.9  \\
\cline{1-4}
Expected update ratio    & 26.6 & 38.7 & 65.0 \\
\Xhline{1pt}
\end{tabular}
\label{tab:accum_exp}
\end{table}

The CMAX-CAMEL engine avoids this problem by using four banks per channel and assigning each bilinear tap to a bank according to the parity bits of its pixel coordinate:
{\small\begin{equation}
b=\{y_0[0],x_0[0]\}.
\end{equation}}%
Because the four neighbors of a bilinear stencil always have distinct least-significant-bit pairs, the four taps of an event are guaranteed to fall into different banks. Each bank stores one even/odd coordinate class, so the bank-local address space becomes $\lceil H_s/2\rceil \times \lceil W_s/2\rceil$. The base address is
{\small{\begin{equation}
a_{00}=\left\lfloor\frac{y_0}{2}\right\rfloor \cdot \left\lceil\frac{W_s}{2}\right\rceil + \left\lfloor\frac{x_0}{2}\right\rfloor,
\end{equation}}}%
and the remaining addresses follow by the corresponding horizontal and vertical increments. Applying this mapping to all four channels yields a $4$-channel $\times$ $4$-bank, 16-lane write organization. In parallel, the bilinear weights derived from $(\alpha_x,\alpha_y)$ are combined with the event polarity $p$ and the Jacobian rows $(\mathbf{r}_x,\mathbf{r}_y)$ to generate per-tap deltas for $\mathrm{IWE}$ and $\mathrm{dIWE}_{x,y,z}$. Each event is therefore converted into 16 tuples of the form $(\text{bank}, \text{address}, \delta)$ together with $(p_{\mathrm{ref}}, p_{\mathrm{act}}, \texttt{last\_in\_pg})$ and forwarded to accumulation. This conflict-free banking removes the main source of write stalls and lowers latency.

\noindent \textit{\textbf{Local Accumulation and Pending Merge.}}
Although banked bilinear voting eliminates structural write conflicts, committing every event contribution to memory would still incur excessive effective memory accesses. 
The CMAX-CAMEL engine therefore applies a two-level reduction scheme: local accumulation followed by pending merge.


The first level exploits the pixel-group locality created by sorting. Events are grouped by the reference index $p_{\mathrm{ref}}$, whereas their actual destination after the current-stage warp is determined by $p_{\mathrm{act}}$ and the corresponding bank/address pair. 
The CMAX-CAMEL engine accumulates only \emph{inlier} events satisfying $p_{\mathrm{act}}=p_{\mathrm{ref}}$ into 16 local registers (4 channels $\times$ 4 taps), while \emph{outlier} events with $p_{\mathrm{act}}\neq p_{\mathrm{ref}}$ are sent to an outlier FIFO. When \texttt{last\_in\_pg} arrives, the accumulated inlier sum is emitted once to an inlier FIFO, transforming event-level memory updates into pixel-group-level partial sums. \refTable{tab:accum_exp} shows that the active pixel-group ratio is much smaller than the total event count, while the outlier ratio remains low, confirming that most events can be absorbed locally within a stage.

The second level, pending merge, further reduces redundancy among the updates that remain after local accumulation. Each lane keeps a pending register containing the most recent outstanding update. A new item is merged immediately if its address matches the pending one; otherwise, the pending value is first committed and then replaced. Under the bilinear $2\times2$ stencil, neighboring pixel groups frequently generate updates to the same bank-local address, and even outliers remain within at most a one-pixel displacement, so the pending-hit probability is high. As shown in \refTable{tab:pend}, the measured reduction exceeds the reduction expected from local accumulation alone, especially at full resolution where spatial overlap becomes more frequent. Together, local accumulation and pending merge convert event-wise updates into a much smaller number of memory update operations, thereby reducing effective memory accesses, energy, and latency.

\begin{table}[t]
\centering
\caption{Memory-update reduction from pending merge.}
\vskip -8pt
\footnotesize
\small
  \renewcommand{\arraystretch}{0.95}
\begin{tabular}{c|c|c|c}
\Xhline{1pt}
\textbf{Metric} &
\textbf{Low-Res} &
\textbf{Mid-Res} &
\textbf{Full-Res} \\
\Xhline{0.8pt}
Expected Reduction (\%) & 72.3 & 60.2 & 31.9 \\
Measured Reduction (\%) & 85.8 & 76.7 & 56.2 \\
\Xhline{1pt}
\end{tabular}
\label{tab:pend}
\end{table}

\noindent \textit{\textbf{Streaming Gaussian Smoothing with On-the-Fly Statistics.}}
The final stage performs Gaussian smoothing on the accumulated IWE and derivative-image channels and directly converts the blurred outputs into the statistics needed for the CMAX objective and gradient. The four channels are processed row-by-row at 2 pixels/clk. To preserve the original Gaussian-smoothed CMAX objective while keeping the hardware compact, the CMAX-CAMEL engine exploits filter separability and implements the 2-D Gaussian filter as a horizontal 1-D FIR followed by a vertical line-buffer stage. The quarter-, half-, and full-resolution stages use 3-, 5-, and 9-tap kernels, respectively. Instead of instantiating separate filters for each stage, the processor is provisioned for the maximum 9-tap case, and lower-resolution stages simply disable the unused taps.


Blurred images are not written back to a separate frame buffer. Instead, the processor updates the final statistics on the fly as each blurred pixel emerges from the filter. Let $P=H_sW_s$ denote the number of pixels at the current stage. The variance objective and its gradient are
{\small{
\begin{equation}
\mathrm{Var}(I_\sigma) = \frac{1}{P}\sum_{\mathbf{x}}\!\left(I_\sigma(\mathbf{x}) - \bar{I}_\sigma\right)^2,\;
\frac{\partial C}{\partial \omega_j} = \frac{2}{P}\sum_{\mathbf{x}}\!\left(I_\sigma(\mathbf{x}) - \bar{I}_\sigma\right)D_{\sigma,j}(\mathbf{x})
\end{equation}
}}%
where {\small $D_{\sigma,j}(\mathbf{x})$} denotes the blurred derivative image corresponding to {\small$\mathrm{dIWE}_j$}. 
Defining
{\small${\small S_1 = \sum_{\mathbf{x}} I_\sigma(\mathbf{x})}$, 
${\small S_2 = \sum_{\mathbf{x}} I_\sigma^2(\mathbf{x})}$,  
${\small G_j = \sum_{\mathbf{x}} I_\sigma(\mathbf{x})D_{\sigma,j}(\mathbf{x})}$, \\
${\small T_j = \sum_{\mathbf{x}} D_{\sigma,j}(\mathbf{x})}$, }
the same quantities can be computed as
{\small
\begin{equation}
\mathrm{Var}(I_\sigma)=\frac{S_2}{P}-\left(\frac{S_1}{P}\right)^2,
\qquad
\frac{\partial C}{\partial \omega_j}
=
\frac{2}{P}\left(G_j-\frac{S_1T_j}{P}\right).
\end{equation}}%
Thus, only running sums are maintained, eliminating any additional image-sized memory after blur. This streaming realization removes an entire writeback/readback pass for blurred images and further reduces effective memory accesses.

Taken together, the proposed architecture separates runtime execution control from the repeated CMAX datapath while preserving the original Gaussian-smoothed CMAX formulation. The controller determines stage transitions, whereas the datapath executes each stage with conflict-free writes, reduced effective memory accesses, and low processing latency. This combination makes runtime-adaptive CMAX practical for always-on, low-power edge deployment.

\section{Implementation and Evaluation}\label{sec:exp}

\subsection{Prototype Processor}\label{sec:setup}

\begin{figure}[t]
\centering
\includegraphics[width=0.85\columnwidth]{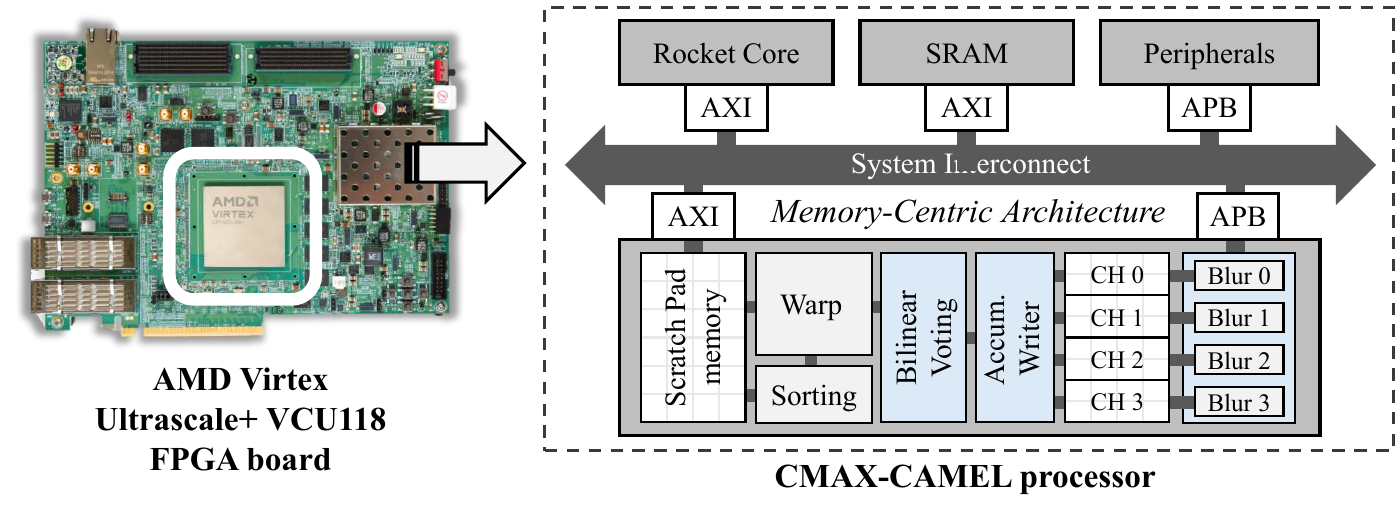}
\vskip -10pt
\caption{CMAX-CAMEL prototype processor used for evaluation.}
\label{fig:fpga}
\end{figure}

\refFigure{fig:fpga} shows the prototype platform used for hardware evaluation. 
The CMAX-CAMEL engine was implemented at RTL in Verilog HDL and integrated into an RISC-V SoC platform comprising a Rocket core~\cite{Rocket}, 128KB SRAM, and a lightweight NoC~\cite{Han:TCAD18}. 
RISC-V eXpress (RVX), an EDA tool widely adopted for developing processors on the RISC-V platform~\cite{Han:IoT2021,Park:TCASI24,Lee:IoTJ25,Cho:DATE26,Min:DAC26}, was used for the SoC integration.
The complete system was prototyped on a Xilinx Virtex UltraScale+ FPGA board~\cite{VCU118} at 200\,MHz. To estimate power and energy beyond the FPGA prototype, the same RTL was synthesized using Synopsys Design Compiler~\cite{DesignCompiler} with the Nangate 45\,nm library~\cite{Nangate45}, and the on-chip memories were modeled as scratchpad RAMs in CACTI~\cite{Balasubramonian:TACO2017} under the same 45\,nm Low-power setting. Dynamic logic power was obtained from synthesis, while dynamic memory energy and leakage were derived from measured access counts and the CACTI models. The on-chip storage was dimensioned for 40,000-event windows and includes the 4-channel$\times$4-bank IWE/dIWE memory array, seven sorting and intermediate-table memories, and 36 line buffers that support up to 9-tap kernels across four blur modules.\refTable{tab:resource} summarizes the FPGA resource usage obtained from Xilinx Vivado~\cite{Vivado} and 45\,nm logic-power estimate of the CMAX-CAMEL processor, while \refTable{tab:mem} reports the representative per-access dynamic energy, total leakage power, and total capacity of each on-chip memory group; for the sorting group, the representative access energy is computed as a stage-weighted average from the measured access distribution.

\begin{table}[t]
   \caption{FPGA resource utilization of the CMAX-CAMEL prototype processor and 45\,nm logic-power estimates.}
   \vskip -8pt
   \centering
   \resizebox{0.95\columnwidth}{!}{
   \renewcommand{\arraystretch}{1.05}
   \begin{tabular}{l|c|c|c|c}
   \Xhline{1pt}
   \multicolumn{1}{c|}{IPs} & {LUTs} & {FFs} & {DSPs} & {Power ($\mathrm{mW}$)} \\ \Xhline{0.8pt}
   \textbf{Full Prototype Processor} & \textbf{66,451} & \textbf{50,575} & \textbf{382} & \textbf{100.35} \\ 
   $\llcorner$ Rocket RISC-V Core      & 15,556 & 9,883  & 4   & 19.30 \\ 
   $\llcorner$ SRAM                    & 163    & 315    & -   & 2.53  \\ 
   $\llcorner$ Peripherals             & 2,761  & 6,014  & -   & 12.66 \\ 
   $\llcorner$ System Interconnect     & 4,934  & 8,103  & -   & 23.09 \\ 
   \rowcolor[HTML]{F2F2F2}
   $\llcorner$ \textbf{CMAX-CAMEL Engine} & \textbf{43,037} & \textbf{26,260} & \textbf{378} & \textbf{42.78} \\ 
   \rowcolor[HTML]{F2F2F2}
   \hspace{0.9em}$\llcorner$ Sorting    & 872    & 1,051  & -   & 0.36  \\ 
   \rowcolor[HTML]{F2F2F2}
   \hspace{0.9em}$\llcorner$ Warp      & 2,373  & 2,847  & 121 & 11.06 \\ 
   \rowcolor[HTML]{F2F2F2}
   \hspace{0.9em}$\llcorner$ Bilinear Voting & 4,881  & 2,377  & 53  & 5.22  \\ 
   \rowcolor[HTML]{F2F2F2}
   \hspace{0.9em}$\llcorner$ Accumulation writer     & 6,513  & 4,549  & -   & 14.70 \\ 
   \rowcolor[HTML]{F2F2F2}
   \hspace{0.9em}$\llcorner$ Blur $\times 4$  & 28,398 & 15,436 & 204 & 11.44 \\
   \Xhline{1pt}
   \end{tabular}
   }
   \label{tab:resource}
\end{table}
\begin{table}[t]
   \caption{Per-access dynamic read/write energy $E_{read}/E_{write}$, total leakage power $P_{total.lkg}$, and memory size of each memory group.}  
   \vskip -8pt
   \centering
   \resizebox{0.85\columnwidth}{!}{
   \renewcommand{\arraystretch}{1.05}
   \begin{tabular}{c|c|c|c}
   \Xhline{1pt}
   \shortstack{\textbf{Memory}\\\textbf{Group}} &
   \shortstack{$\bm{E_{read}}$ / $\bm{E_{write}}$\\\textbf{(pJ/access)}} &
   \shortstack{$\bm{P_{total.lkg}}$\\\textbf{(mW)}} &
   \shortstack{\textbf{Memory Size}\\\textbf{(KB)}} \\
   \Xhline{0.8pt}
   IWE~/~dIWE$_{xyz}$ & 11.26 / 8.07  & 12.39 & 675 \\
   Raw events            & 22.66 / 21.44 & 3.08  & 156 \\
   Sorting buffer         & 9.71 / 8.19   & 10.19 & 520 \\
   line-buffer & 9.18 / 7.83   & 1.43  & 68 \\
   \Xhline{1pt}
   \end{tabular}
   }
   \label{tab:mem}
\end{table}


To isolate the benefit of the proposed architecture, we also implemented a baseline prototype processor on the same platform. The baseline uses the same coarse-to-fine execution structure and the same stage thresholds $\{\tau_s\}$ as CMAX-CAMEL, but removes the two key memory-centric mechanisms: conflict-free banked write mapping and local accumulation. Because sorting provides little benefit beyond subsampling without local accumulation, the baseline skips sorting at the full-resolution stage. 
Hardware evaluation is conducted using fixed 40,000-event windows from the \texttt{poster} sequence of the Event Camera Dataset and Simulator. The minimum window duration in this sequence is 5.72,ms; therefore, this value is used as the real-time design criterion, and an implementation is regarded as real-time capable if its average processing latency does not exceed this bound.

At runtime, DVS events are assumed to be written directly into on-chip event memory through a dedicated sensor interface. The Rocket core repeatedly reads the variance and gradient produced by the engine, updates $\boldsymbol{\omega}$ using the CG-PR~\cite{Polak:CGPR69} optimizer, and controls stage transitions according to Algorithm~\ref{alg:runtime_stage}. Latency and energy are measured from the moment an event window becomes available in on-chip memory until optimization for that window completes, and all reported results are averaged over the full set of windows.


\subsection{Evaluation Results}\label{sec:setup}

Compared with the baseline, the CMAX-CAMEL processor reduces total on-chip memory accesses by 41.95\%, primarily because local accumulation and pending merge absorb repeated updates within and across neighboring pixel groups. In addition, the conflict-free banked-write organization removes the write serialization caused by bilinear voting. As a result, the average window processing latency is reduced by 53.27\%. Although the baseline uses the same coarse-to-fine execution policy, its unresolved memory bottlenecks prevent it from meeting the real-time design criterion, whereas CMAX-CAMEL satisfies this criterion by reducing the average latency below the required bound.

\refTable{tab:energy_breakdown} reports the corresponding energy breakdown. Relative to the baseline, CMAX-CAMEL reduces memory read/write energy by 67.18\%, logic plus memory-leakage energy by 47.38\%, and total energy by 52.17\%. The larger reduction in memory read/write energy than in raw access count reflects the fact that dynamic memory energy depends not only on the number of accesses but also on the accessed memory group and the read/write mix. Of the total energy reduction, 31.15\% comes from the direct reduction in memory-access activity, while the remaining 68.85\% comes from the shorter processing time, which proportionally lowers both logic and leakage energy.

Finally, we verify that the processor implementation preserves algorithmic accuracy. On the \texttt{poster} sequence, the software implementation of the runtime-adaptive method achieves an RMSE of 9.56, whereas the quantized hardware implementation on the CMAX-CAMEL processor achieves 9.61. This small difference shows that the proposed architecture and hardware quantization preserve the behavior of the runtime-adaptive algorithm while delivering substantial reductions in memory accesses, latency, and energy.

\begin{table}[t]
\caption{Comparison of memory read/write energy $E_{mem.R/W}$, the sum of memory leakage energy $E_{mem.lkg}$ and logic energy $E_{logic}$, and total energy $E_{total}$ between CMAX-CAMEL and the baseline hardware on the \texttt{poster} sequence, together with the corresponding energy savings, where energy values are reported in $\bm{\mu}$J.}
   \vskip -8pt
   \centering
   \resizebox{0.94\columnwidth}{!}{
   \renewcommand{\arraystretch}{1.08}
   \begin{tabular}{c|c|c|c}
   \Xhline{1pt}
   {\bfseries Design} &
   \shortstack{\bfseries $\bm{E_{mem.R/W}}$} &
   \shortstack{\bfseries $\bm{E_{logic}+E_{mem.lkg}}$} &
   \shortstack{\bfseries $\bm{E_{total}}$} \\
   \Xhline{0.8pt}
   Baseline HW         & 339.80 & 1064.77 & 1404.57 \\
   \textbf{CMAX-CAMEL} & \textbf{111.53} & \textbf{560.27} & \textbf{671.80} \\
   \hline
   Energy Savings      & -67.18\% & -47.38\% & -52.17\% \\
   \Xhline{1pt}
   \end{tabular}
   }
   \label{tab:energy_breakdown}
\end{table}

\section{Conclusion}
This work presented CMAX-CAMEL, a coarse-to-fine adaptive, memory-efficient, and low-power edge processor for contrast maximization in event-based motion estimation. CMAX-CAMEL combines a runtime-adaptive execution policy with a memory-centric engine architecture that reduces effective memory accesses, alleviates write conflicts, and lowers processing latency while preserving the original Gaussian-smoothed CMAX formulation. Implemented on a Virtex FPGA prototype and evaluated with 45\,nm power models, the proposed processor achieved up to 19\% higher estimation accuracy than fixed coarse-to-fine execution, reduced processing latency by 53.3\%, lowered effective memory accesses by 42\%, and reduced total energy by 52.2\%. These results show that runtime-adaptive control and memory-centric datapath design together provide an effective processor-level solution for real-time, low-power edge deployment of CMAX.

\bibliographystyle{unsrt}
\bibliography{reference}

@article{CHOI:AEJ25,
title = {{E-BTS}: A low-power Event-driven Blink Tracking System with hardware-software co-optimized design for real-time driver drowsiness detection},
journal = {Alexandria Engineering Journal},
volume = {128},
pages = {867-877},
year = {2025},
author = {Jongin Choi and Eunjin Choi and Seonghyun Choi and Woojoo Lee},
}

@ARTICLE{Han:IoT2021,
	author={Han, Kyuseung and Lee, Sukho and Oh, Kwang-Il and Bae, Younghwan and Jang, Hyeonguk and Lee, Jae-Jin and Lee, Woojoo and Pedram, Massoud},
	journal={IEEE Internet of Things Journal},
	title={Developing {TEI}-Aware Ultralow-Power SoC Platforms for IoT End Nodes},
	year={2021},
	volume={8},
	number={6},
	pages={4642-4656}
}

@article{Min:DAC26,
      title={{TT-SEAL: TTD-Aware Selective Encryption for Adversarially-Robust and Low-Latency Edge AI}}, 
      author={Kyeongpil Min and Sangmin Jeon and Jae-Jin Lee and Woojoo Lee},
      year={2026},
      journal={https://arxiv.org/abs/2602.22238}, 
}

@ARTICLE{Park:TCASI24,
  author={Park, Jina and Han, Kyuseung and Choi, Eunjin and Lee, Jae-Jin and Lee, Kyeongwon and Lee, Woojoo and Pedram, Massoud},
  journal={IEEE Transactions on Circuits and Systems I: Regular Papers}, 
  title={Designing Low-Power {RISC-V} Multicore Processors With a Shared Lightweight Floating Point Unit for {IoT} Endnodes}, 
  year={2024},
  volume={71},
  number={9},
  pages={4106-4119},
}

@ARTICLE{Lee:IoTJ25,
  author={Lee, Kyeongwon and Jeon, Sangmin and Lee, Kangju and Lee, Woojoo and Pedram, Massoud},
  journal={IEEE Internet of Things Journal}, 
  title={Radar-{PIM}: Developing {IoT} Processors Utilizing {Processing-in-Memory} Architecture for Ultrawideband-Radar-Based Respiration Detection}, 
  year={2025},
  volume={12},
  number={1},
  pages={515-530},
}

@ARTICLE{Han:TCAD18,
  author={Han, Kyuseung and Lee, Jae-Jin and Lee, Jinho and Lee, Woojoo and Pedram, Massoud},
  journal={IEEE Transactions on Computer-Aided Design of Integrated Circuits and Systems}, 
  title={TEI-NoC: Optimizing Ultralow Power NoCs Exploiting the Temperature Effect Inversion}, 
  year={2018},
  volume={37},
  number={2},
  pages={458-471},
  doi={10.1109/TCAD.2017.2693269}}

@INPROCEEDINGS{Cho:DATE26,
  author={Cho, Eun-Su and Choi, Jongin and Jin, Jeongmin and Lee, Jae-Jin and Lee, Woojoo},
  booktitle={2026 Design, Automation \& Test in Europe Conference (DATE)}, 
  title={{FiCABU: A Fisher-Based, Context-Adaptive Machine Unlearning Processor for Edge AI}}, 
  year={2026},
  volume={},
  number={},
  pages={1-7}
  }

@article{gehrig:nature2024,
  title={Low-latency automotive vision with event cameras},
  author={Gehrig, Daniel and Scaramuzza, Davide},
  journal={Nature},
  volume={629},
  number={8014},
  pages={1034--1040},
  year={2024},
  publisher={Nature Publishing Group UK London}
}

@article{kuhne:sensors2024,
  title={Low latency visual inertial odometry with on-sensor accelerated optical flow for resource-constrained {UAVs}},
  author={K{\"u}hne, Jonas and Magno, Michele and Benini, Luca},
  journal={IEEE Sensors Journal},
  volume={25},
  number={5},
  pages={7838--7847},
  year={2024},
  publisher={IEEE}
}

@article{wang:TIM2024,
  title={A survey of visual {SLAM} in dynamic environment: The evolution from geometric to semantic approaches},
  author={Wang, Yanan and Tian, Yaobin and Chen, Jiawei and Xu, Kun and Ding, Xilun},
  journal={IEEE Transactions on Instrumentation and Measurement},
  volume={73},
  pages={1--21},
  year={2024},
  publisher={IEEE}
}

@article{xu:robotics2025,
  title={Airslam: An efficient and illumination-robust point-line visual slam system},
  author={Xu, Kuan and Hao, Yuefan and Yuan, Shenghai and Wang, Chen and Xie, Lihua},
  journal={IEEE Transactions on Robotics},
  volume={41},
  pages={1673--1692},
  year={2025},
  publisher={IEEE}
}

@ARTICLE{lichtsteiner:JSSC2008,
  author={Lichtsteiner, Patrick and Posch, Christoph and Delbruck, Tobi},
  journal={IEEE Journal of Solid-State Circuits}, 
  title={A 128$\times$ 128 120 {dB} 15 $\mu$s {Latency Asynchronous Temporal Contrast Vision Sensor}}, 
  year={2008},
  volume={43},
  number={2},
  pages={566-576},
  doi={10.1109/JSSC.2007.914337}}

@ARTICLE{Rebecq:LRA17,
  author={Rebecq, Henri and Horstschaefer, Timo and Gallego, Guillermo and Scaramuzza, Davide},
  journal={IEEE Robotics and Automation Letters}, 
  title={{EVO: A Geometric Approach to Event-Based 6-DOF Parallel Tracking and Mapping in Real Time}}, 
  year={2017},
  volume={2},
  number={2},
  pages={593-600},
  doi={10.1109/LRA.2016.2645143}
}

@Article{Cimarelli:sensors25,
AUTHOR = {Cimarelli, Claudio and Millan-Romera, Jose Andres and Voos, Holger and Sanchez-Lopez, Jose Luis},
TITLE = {{Hardware, Algorithms, and Applications of the Neuromorphic Vision Sensor: A Review}},
JOURNAL = {Sensors},
VOLUME = {25},
YEAR = {2025},
NUMBER = {19},
ARTICLE-NUMBER = {6208},
URL = {https://www.mdpi.com/1424-8220/25/19/6208},
PubMedID = {41095030},
ISSN = {1424-8220},
}

@article{aliakbarpour:Sensors2024,
  title={Emerging trends and applications of neuromorphic dynamic vision sensors: {A} survey},
  author={AliAkbarpour, Hadi and Moori, Ahmad and Khorramdel, Javad and Blasch, Erik and Tahri, Omar},
  journal={IEEE Sensors Reviews},
  volume={1},
  pages={14--63},
  year={2024},
  publisher={IEEE}
}

@article{gallego:TPAMI2020,
  title={{Event-based vision: A survey}},
  author={Gallego, Guillermo and Delbr{\"u}ck, Tobi and Orchard, Garrick and Bartolozzi, Chiara and Taba, Brian and Censi, Andrea and Leutenegger, Stefan and Davison, Andrew J and Conradt, J{\"o}rg and Daniilidis, Kostas and others},
  journal={IEEE transactions on pattern analysis and machine intelligence},
  volume={44},
  number={1},
  pages={154--180},
  year={2020},
  publisher={IEEE}
}

@INPROCEEDINGS{Kueng:IROS16,
  author={Kueng, Beat and Mueggler, Elias and Gallego, Guillermo and Scaramuzza, Davide},
  booktitle={2016 IEEE/RSJ International Conference on Intelligent Robots and Systems (IROS)}, 
  title={Low-latency visual odometry using event-based feature tracks}, 
  year={2016},
  volume={},
  number={},
  pages={16-23},
  doi={10.1109/IROS.2016.7758089}
}

@article{chamorro:robotics2022,
  title={Event-based line {SLAM} in real-time},
  author={Chamorro, William and Sola, Joan and Andrade-Cetto, Juan},
  journal={IEEE Robotics and Automation Letters},
  volume={7},
  number={3},
  pages={8146--8153},
  year={2022},
  publisher={IEEE}
}

@article{zhou:robotics2021,
  title={Event-based stereo visual odometry},
  author={Zhou, Yi and Gallego, Guillermo and Shen, Shaojie},
  journal={IEEE Transactions on Robotics},
  volume={37},
  number={5},
  pages={1433--1450},
  year={2021},
  publisher={IEEE}
}

@ARTICLE{Gallego:LRA17,
  author={Gallego, Guillermo and Scaramuzza, Davide},
  journal={IEEE Robotics and Automation Letters}, 
  title={{Accurate Angular Velocity Estimation With an Event Camera}}, 
  year={2017},
  volume={2},
  number={2},
  pages={632-639},
}

@inproceedings{gallego:CVPR2018,
  title={A unifying contrast maximization framework for event cameras, with applications to motion, depth, and optical flow estimation},
  author={Gallego, Guillermo and Rebecq, Henri and Scaramuzza, Davide},
  booktitle={Proceedings of the IEEE conference on computer vision and pattern recognition},
  pages={3867--3876},
  year={2018}
}

@ARTICLE{Kim:LRA21,
  author={Kim, Haram and Kim, H. Jin},
  journal={IEEE Robotics and Automation Letters}, 
  title={{Real-Time Rotational Motion Estimation With Contrast Maximization Over Globally Aligned Events}}, 
  year={2021},
  volume={6},
  number={3},
  pages={6016-6023},
  doi={10.1109/LRA.2021.3088793}
}

@ARTICLE{Zhou:TNNLS23,
  author={Zhou, Yi and Gallego, Guillermo and Lu, Xiuyuan and Liu, Siqi and Shen, Shaojie},
  journal={IEEE Transactions on Neural Networks and Learning Systems}, 
  title={{Event-Based Motion Segmentation With Spatio-Temporal Graph Cuts}}, 
  year={2023},
  volume={34},
  number={8},
  pages={4868-4880},
}

@ARTICLE{Shiba:TPAMI24,
  author={Shiba, Shintaro and Klose, Yannick and Aoki, Yoshimitsu and Gallego, Guillermo},
  journal={IEEE Transactions on Pattern Analysis and Machine Intelligence}, 
  title={{Secrets of Event-Based Optical Flow, Depth and Ego-Motion Estimation by Contrast Maximization}}, 
  year={2024},
  volume={46},
  number={12},
  pages={7742-7759},
}

@article{guo:Robotics2024,
  title={{CMax-SLAM}: Event-based rotational-motion bundle adjustment and SLAM system using contrast maximization},
  author={Guo, Shuang and Gallego, Guillermo},
  journal={IEEE Transactions on Robotics},
  volume={40},
  pages={2442--2461},
  year={2024},
  publisher={IEEE}
}

@inproceedings{stoffregen:ICCV2019,
  title={Event-based motion segmentation by motion compensation},
  author={Stoffregen, Timo and Gallego, Guillermo and Drummond, Tom and Kleeman, Lindsay and Scaramuzza, Davide},
  booktitle={Proceedings of the IEEE/CVF International Conference on Computer Vision},
  pages={7244--7253},
  year={2019}
}

@InProceedings{Yamaki:CVPRW25,
  author    = {Yamaki, Ryo and Shiba, Shintaro and Guillermo, Gallego and Aoki, Yoshimitsu},
  title     = {{Iterative Event-based Motion Segmentation by Variational Contrast Maximization}},
  booktitle = {IEEE/CVF Conference on Computer Vision and Pattern Recognition (CVPR) Workshops},
  month     = {June},
  year      = {2025},
  pages     = {4957-4966}
}

@inproceedings{gallego:CVPR2019,
  author    = {Gallego, Guillermo and Gehrig, Mathias and Scaramuzza, Davide},
  title     = {{Focus Is All You Need: Loss Functions for Event-Based Vision}},
  booktitle = {Proc. IEEE/CVF Conf. Comput. Vis. Pattern Recognit. (CVPR)},
  pages     = {12272--12281},
  year      = {2019},
  doi       = {10.1109/CVPR.2019.01256}
}

@inproceedings{wang2025frme,
  title={{FRME: An FPGA-Accelerated Event-Based Real-Time Rotational Motion Estimator for SLAM}},
  author={Wang, Runhua and Zhang, Shen and Yan, Guangyao and Liu, Tianhang and Zhou, Zhiqi and Guo, Shuang and Wei, Boyi and Shi, Chenyang and Wang, Hui and Ha, Yajun},
  booktitle={2025 International Conference on Field Programmable Technology (ICFPT)},
  pages={128--136},
  year={2025},
  organization={IEEE}
}

@article{min2026coarse,
  title={{Coarse-to-Fine Contrast Maximization for Energy-Efficient Motion Estimation in Edge-Deployed Event-Based SLAM}},
  author={Min, Kyeongpil and Choi, Jongin and Lee, Woojoo},
  journal={Micromachines},
  volume={17},
  number={2},
  pages={176},
  year={2026},
  publisher={MDPI}
}

@article{silvano:survey2025,
  title={A survey on deep learning hardware accelerators for heterogeneous hpc platforms},
  author={Silvano, Cristina and Ielmini, Daniele and Ferrandi, Fabrizio and Fiorin, Leandro and Curzel, Serena and Benini, Luca and Conti, Francesco and Garofalo, Angelo and Zambelli, Cristian and Calore, Enrico and others},
  journal={ACM Computing Surveys},
  volume={57},
  number={11},
  pages={1--39},
  year={2025},
  publisher={ACM New York, NY}
}

@article{akkad:TAI2023,
  title={Embedded deep learning accelerators: {A} survey on recent advances},
  author={Akkad, Ghattas and Mansour, Ali and Inaty, Elie},
  journal={IEEE Transactions on Artificial Intelligence},
  volume={5},
  number={5},
  pages={1954--1972},
  year={2023},
  publisher={IEEE}
}

@article{golpayegani:TAAS2024,
  title={Adaptation in edge computing: a review on design principles and research challenges},
  author={Golpayegani, Fateneh and Chen, Nanxi and Afraz, Nima and Gyamfi, Eric and Malekjafarian, Abdollah and Sch{\"a}fer, Dominik and Krupitzer, Christian},
  journal={ACM Transactions on Autonomous and Adaptive Systems},
  volume={19},
  number={3},
  pages={1--43},
  year={2024},
  publisher={ACM New York, NY}
}

@article{tuli:JNPA2023,
  title={{AI augmented Edge and Fog computing: Trends and challenges}},
  author={Tuli, Shreshth and Mirhakimi, Fatemeh and Pallewatta, Samodha and Zawad, Syed and Casale, Giuliano and Javadi, Bahman and Yan, Feng and Buyya, Rajkumar and Jennings, Nicholas R},
  journal={Journal of Network and Computer Applications},
  volume={216},
  pages={103648},
  year={2023},
  publisher={Elsevier}
}

@article{Mueggler:IJRR17,
  title={The event-camera dataset and simulator: Event-based data for pose estimation, visual odometry, and {SLAM}},
  author={Mueggler, Elias and Rebecq, Henri and Gallego, Guillermo and Delbruck, Tobi and Scaramuzza, Davide},
  journal={The International journal of robotics research},
  volume={36},
  number={2},
  pages={142--149},
  year={2017},
  publisher={SAGE Publications Sage UK: London, England}
}

@MISC{Rocket,
  author       = {SiFIVE},
  howpublished = {\href{https://github.com/chipsalliance/rocket-chip}{\nolinkurl{https://github.com/chipsalliance/rocket-chip}}},
  note         = {Accessed \TODAY}
}

@MISC{VCU118,
author       = {AMD},
  title       = {{VCU118}},
  howpublished = {\href{https://www.amd.com/en/products/adaptive-socs-and-fpgas/evaluation-boards/vcu118.html}{\nolinkurl{https://www.amd.com/en/products/adaptive-socs-and-fpgas/evaluation-boards/vcu118.html}}},
  note         = {Accessed \TODAY}
}

@MISC{DesignCompiler,
  author       = {Synopsys},
   title        = {{DesignCompiler}},
  howpublished = {\href{https://www.synopsys.com/implementation-and-signoff/rtl-synthesis-test/dc-ultra.html}{\nolinkurl{https://www.synopsys.com/implementation-and-signoff/rtl-synthesis-test/dc-ultra.html}}},
  note         = {Accessed \TODAY}
}

@misc{Nangate45,
  author       = {{Nangate Inc.}},
  title        = {{45nm Open Cell Library}},
  howpublished = {\url{https://eda.ncsu.edu/freepdk/freepdk45}},
  note         = {Accessed March 2026}
}

@article{Balasubramonian:TACO2017,
  title={{CACTI} 7: New tools for interconnect exploration in innovative off-chip memories},
  author={Balasubramonian, Rajeev and Kahng, Andrew B and Muralimanohar, Naveen and Shafiee, Ali and Srinivas, Vaishnav},
  journal={ACM Transactions on Architecture and Code Optimization (TACO)},
  volume={14},
  number={2},
  pages={1--25},
  year={2017},
  publisher={ACM New York, NY, USA}
}

@MISC{Vivado,
	author={Xilinx},
	title = {{Vivado}},
	howpublished = {\url{https://www.xilinx.com/support/download.html}},
	note = {Accessed \today}
}

@article{Polak:CGPR69,
  title={Note sur la convergence de m\'ethodes de directions conjugu\'ees},
  author={Polak, E. and Ribière, G.},
  journal={Revue fran\c{c}aise d'informatique et de recherche op\'erationnelle. S\'erie rouge},
  volume={3},
  number={R1},
  pages={35--43},
  year={1969},
  publisher={Dunod}
}
\end{document}